# SL—a "quick and dirty" but working intermediate language for SVP systems

Raphael 'kena' Poss
University of Amsterdam, The Netherlands
`r.c.poss@uva.nl`

June 8, 2018

**Abstract**

The CSA group at the University of Amsterdam has developed SVP, a framework to manage and program many-core and hardware multi-threaded processors. In this article, we introduce the intermediate language SL, a common vehicle to program SVP platforms. SL is designed as an extension to the standard C language (ISO C99/C11). It includes primitive constructs to bulk create threads, bulk synchronize on termination of threads, and communicate using word-sized dataflow channels between threads. It is intended for use as target language for higher-level parallelizing compilers. SL is a research vehicle; as of this writing, it is the only interface language to program a main SVP platform, the new Microgrid chip architecture. This article provides an overview of the language, to complement a detailed specification available separately.

# Contents

# 1 Introduction

Multi-core processors are a stepping stone towards the integration of entire computing systems on chip, which is the current trend for chip design as long as Moore's law continues to hold. To manage the complexity of large many-core processor chips, with and without hardware multithreading, while enabling their expected benefits for performance and execution efficiency, the Computer Systems Architecture research group[1] at the University of Amsterdam is developing a framework called the System Virtualization Platform[2] (SVP).

The essence of SVP is to capture concurrency management—thread creation, synchronization and communication—at the lowest level of the software stack, i.e. even below the level of operating systems, and subsequently offer a uniform interface to software for concurrency control (cf. section 3). SVP has been implemented in software emulations on top of POSIX platforms [29], but also as a hardware protocol in an experimental hardware multithreaded, many-core architecture called the Microgrid [1, 8, 7, 24].

Any implementation of SVP, including Microgrid platforms in particular, offers low-level primitives for concurrency creation, synchronization, the definition of dataflow communication channels, and communication over these channels. However, the precise implementation and interface semantics may differ slightly depending on the concrete SVP platform. To capture the semantics of SVP in a *common language* that can target all envisioned SVP platforms, two extensions to the C language were devised: first $\mu$TC, then SL.

The language $\mu$TC was historically the first developed [17, 4, 2, 5]. However, $\mu$TC eventually proved impractical: we were able to show [18, App. G] that no compiler could be possibly be built that could translate $\mu$TC to one of the envisioned SVP platforms, the UTLEON3 single-core design on FPGA [11]. Instead, we designed and implemented SL, another C extension inspired from $\mu$TC able to program all SVP platforms produced so far.

This article provides an introduction to SL, organized as follows. In section 2, we provide a first example SL program for the impatient reader. In section 3 we provide a short introduction to SVP. In section 4 we introduce the language specification and provide a high-level overview of its main constructs. In section 5 we expose our design rationale and an introduction to the SL compilation tool chain. In section 6 we review ongoing and planned improvement efforts. We conclude and summarize in section 7.

# 2 "Hello world" in SL

```
#include <stdio.h>
int main(void) {
    printf("hello_world\n");
    return 0;
}
```
Listing 1: "Hello world" in SL.

---

[1] http://csa.science.uva.nl/
[2] http://www.svp-home.org/

As a C extension, SL supports most programs that are also valid in C. In particular, the typical "hello world" program remains unchanged in SL, as given in listing 1.

```
#include <stdio.h>
// This thread program ``foo'' increments the value it
// receives over its input channel, propagates the
// incremented value to its output channel, then prints
// a digit on the C standard output.
sl_def(foo, , sl_shparm(int, a)) {
    sl_setp(a, sl_getp(a) + 1);
    printf("%d", sl_getp(a));
} sl_enddef

int main(void) {
    // Create a dependent family of 10 threads.
    sl_create(,, 0, 10, 1, 0,, foo, sl_sharg(int, x));
    // Push 0 as source value for the dataflow channel.
    sl_seta(x, 0);
    // Wait on termination of the family.
    sl_sync();
    // After the family terminates, the channel endpoint
    // reveals the final value.
    printf("%d\n", sl_geta(x));
    return 0;
}
```

Listing 2: Listing for `10threads.c`.

A more interesting example is given in listing 2: this program prints the ten decimal digits to standard output, then the value "10" in decimal, then a newline character. Moreover, each call to the C library to print a digit is performed *concurrently* by a separate thread: the order in which each digit is printed is non-deterministic. For example, both the outputs "012345678910" and "430729861510" are valid. This program demonstrates the following SL constructs:

- `sl_def`...`sl_enddef` defines a *thread function*, which is a special kind of C function that can be run by bulk created threads.
- `sl_shparm` defines a pair of dataflow channels provided as input/output to a thread function.
- `sl_getp`/`sl_setp` are read/write operations on dataflow channels defined with `sl_shparm`.
- `sl_create`...`sl_sync` defines a concurrent *family* of multiple bulk created threads. The positional arguments include the *index range* (defines how many threads are created in total) and *window size* (how many threads are created simultaneously).
- `sl_sharg` defines a pair of dataflow endpoints to communicate with a family from its creator thread. The data type is explicit and must match the channel type in the corresponding `sl_def` construct.
- `sl_geta`/`sl_seta` are read/write operations on the channel endpoints defined by `sl_sharg`.

| Event name | Description | Parameters |
|---|---|---|
| allocate | Context allocation | resource identifier, desired context type, failure mode |
| configure | Context configuration | context identifier, channel interface, logical thread index range, window size |
| create | Bulk creation | context identifier, thread function address |
| sync | Bulk synchronization | context identifier |
| release | Context de-allocation | context identifier |
| read/write | Local channel access | relative channel identifier |
| put/get | Remote channel access | context identifier, relative channel identifier |

Table 1: SVP control events.

# 3 Introduction to SVP primitives

The SVP model captures concurrency as follows:
- each thread of execution belongs to a *family* of *logical threads*;
- entire families of multiple logical threads execute within a *context* that must be reserved at a named *resource* (location) on chip, e.g. a cluster of cores;
- once its context is reserved, a family of multiple logical threads is *created* at this context using a single *bulk creation* operation;
- the number of logical threads is specified by *configuring* a context prior to bulk creation;
- the threads belonging to families created in separate resources are scheduled independently;
- within a family, a *window size* parameter limits how many hardware threads are allocated to the family on each core (0 indicates no limit), and thus determines how many logical threads execute simultaneously; any excess logical threads are serialized over the allocated hardware threads, i.e. a new logical thread is created only once the previous logical thread has terminated;
- a thread can *bulk synchronize* on a family running at a given context, i.e. wait until all logical threads in the family have terminated;
- while a family is running, a thread outside of the family can communicate using *remote channel accesses* with the family's threads;
- at least two communication patterns are supported: *global* channels which implement a single producer (outside of the family) and multiple consumers (each logical thread within the family); and *shared* channels which implement daisy chained (thread-to-thread in range order) communication within the family.

The corresponding SVP control events are summarized in table 1. They are implemented as ISA machine instructions on the Microgrid architecture [18, Chap. 2&3, App. D], and as a library API in a POSIX thread emulation [29].

# 4 Specification and overview

A full specification of the SL core language has been published in [18, App. I]; it is formulated as fully-fledged *addendum* to the ISO C standards [15, 16]. Of course, while necessary, a detailed language specification like ISO C or derived documents is practically useless to provide an introduction to newcomers. The rest of this section thus complements the specification with a high-level overview.

## 4.1 Thread functions

```
sl_def( F [, , thread-parameters...] )
{
    /* body... */
}
sl_enddef
```
Listing 3: SL construct to define thread functions.

Logical threads on the SVP platform run *thread functions*. Thread functions are defined using the `sl_def` in SL, as depicted in listing 3.

This construct defines the thread function named "F." The subsequent comma-separated list of *thread parameters* defines the communication channel endpoints visible from F. A possible channel endpoint declaration is the construct:

```
sl_glparm( T, P )
```

which defines a channel endpoint named "A" able to transport values of type "T." From this point, any expression of the form:

```
sl_getp( P )
```

in the function body reads from the channel endpoint named "P" and evaluates to an expression of type "T."

Meanwhile, within the thread function body, the declaration:

```
sl_index( I );
```

defines a variable named "I" which will automatically receive, at run-time, the logical index of the thread currently executing the thread function. It has a signed integer type.

```
sl_def(scal, , sl_glparm(int*, a), sl_glparm(int, c))
{
    sl_index(i);
    int *a = sl_getp(a);
    a[i] = a[i] * sl_getp(c);
}
sl_enddef
```
Listing 4: Vector scaling function in SL.

For example, listing 4 implements a scaling function over the elements of an array. It defines two channel endpoints, one to receive the base pointer to an array and another to receive the scaling constant. It then uses its logical index to scale the vector item at that position.

```
sl_def(innerprod, , sl_glparm(int*, a), sl_glparm(int*, b),
       sl_shparm(int, s))
{
  sl_index(i);
  int *a = sl_getp(a), *b = sl_getp(b);
  sl_setp(s, sl_getp(s) + a[i] * b[i]);
}
sl_enddef
```

Listing 5: Partial inner product in SL.

Next to "global" channel endpoints, declared by `sl_glparm`, a thread function can also declare "shared" endpoints with `sl_shparm` which uses otherwise the same syntax. The difference between "global" and "shared" channels was given in section 3. For example, listing 5 implements a partial vector summation. It defines three channel endpoints, two for "global" channels to receive the array bases, and one for a "shared" channel for the partial sum.

For completeness with the C language, it is also possible to define a thread function with the specifier "`sl__static`" with the same visibility rules as `static` in C. Forward or external declarations can also be expressed using `sl_decl`, with the same syntax as `sl_def` but without the function body. For example, the following forward declares a static (local) thread function `foo` with one channel endpoint:

```
sl_decl(foo, sl__static, sl_glarg(int, x));
```

```
sl_def(sscal, , sl_glparm(float*, a), sl_glfparm(float, c))
{
  sl_index(i);
  float *a = sl_getp(a);
  a[i] = a[i] * sl_getp(c);
}
sl_enddef
```

Listing 6: Vector scaling function in SL, using floating point values.

Finally, as an idiosyncrasy of SL motivated by its implementation strategy (cf. section 5.3), the current language definition requires a separate syntax for floating-point (FP) channel endpoints. These must be declared using `sl_glfparm` and `sl_shfparm`. The reason for this separation is that the SL compiler must generate different code for FP channels, and the context-free substitution of SL constructs cannot properly determine whether a given channel type is actually FP. Note however that base pointers to FP arrays are really integer scalars for the purpose of channel declarations. For example, listing 6 updates listing 4 with support for FP values.

```
c-external-declaration:
    ...
    thread-function-definition
    thread-function-declaration
thread-function-declaration:
    sl_decl( identifier [ , [ tfun-spec ] [ , param-spec ]* ] );
thread-function-definition:
    sl_def( identifier [ , [ tfun-spec ] [ , chan-spec ]* ] ) ¬
        c-compound-statement ¬
        sl_enddef
tfun-spec:
    sl__static
chan-spec:
    sl_glparm( endpoint-def )
    sl_shparm( endpoint-def )
    sl_glfparm( endpoint-def )
    sl_shfparm( endpoint-def )
endpoint-def:
    c-declaration-specifiers , identifier
c-declaration:
    ...
    sl_index( identifier );
c-primary-expression:
    ...
    sl_getp( identifier )
c-statement:
    ...
    sl_setp( identifier , c-assignment-expression );
```

Figure 1: BNF grammar for thread function declarations and definitions in SL.

For completeness, we detail the grammar of thread function declarations and definitions in fig. 1.

## 4.2 Family creation

```
1 sl_create ( , , [ start ] , [ limit ] , [ step ] , [ ws ] , , F [ , ... ] );
2 ...
3 sl_sync ();
```

Listing 7: SL construct for family creation and synchronization.

Any thread can itself *create* whole families of logical threads at once and then *synchronize* on their termination. A single `sl_create`...`sl_sync` construct expresses both bulk creation and synchronization, given in listing 7.

This construct expresses the creation of a family of logical threads running the thread function "F." The `start`, `limit` and `step` integer expressions define the *logical range*, while `ws` defines the *window size*. We presented both concepts above in section 3. All four expressions are optional, and default to 0, 1, 1, 0 respectively if left unspecified. Therefore, the program in listing 8 prints "hello world" exactly once.

```
#include <stdio.h>

sl_def(hw) { printf("hello, world\n"); } sl_enddef

int main(void) {
  sl_create(,,,,,,, hw);
  sl_sync();
  return 0;
}
```
Listing 8: "Hello world" using a logical thread in SL.

Note that despite their syntax, the two parts of `sl_create`...`sl_sync` are not two separate C statements. Instead, they are syntactically bound and must appear *within* a compound statement (block enclosed by curly brackets). For example, the syntax in listing 9 is invalid because the `if` construct expects a single statement, whereas listing 10 is valid.

```
if (condition)
    sl_create(...); ... sl_sync();
```
Listing 9: Invalid use of `sl_create`...`sl_sync` as statement.

```
if (condition)
{ sl_create(...); ... sl_sync(); }
```
Listing 10: Valid use of `sl_create`...`sl_sync` as block item.

```
int main(void) {
  float v[5] = { 1, 2, 3, 4, 5 };
  sl_create(,,, 5,,,, sscal,
            sl_glarg(float*,cv), sl_glfarg(float,cc));
  sl_seta(cv, v);
  sl_seta(cc, 3.0);
  sl_sync();
  printf("%f\n", v[2]); // prints 9.0
  return 0;
}
```
Listing 11: Example test program for "sscal."

When the thread function defines channel endpoints, these are connected to the creating thread using `sl_*arg`, as illustrated in listing 11. This example defines a vector of 5 elements and then creates a family of 5 threads running the the "sscal" function defined in listing 6. The channel endpoints in "main" are labeled "cv" and "cc" by `sl_glarg` and `sl_glfarg`. "main" then sends the vector base address and a FP constant as source value for the two global channels to the family using `sl_seta`, to be received by all 5 logical threads using `sl_getp`.

```
1    sl_create(,,, 5,,,, sscal,
2             sl_glarg(float*,,v), sl_glfarg(float,,3.0));
3    sl_sync();
```
Listing 12: Alternate creation of a family running "sscal," using anonymous channels.

When the source values are known at the point `sl_create` is reached, it is possible to combine the definition of channel endpoints with sending a value, as shown in listing 12. This example demonstrates both how to send a value directly from the `sl_create` construct, and the option to leave channel endpoints unnamed.

```
1    int main(void) {
2      int v1[5] = { 1, 2, 3, 4, 5 }, v2[5] = { 3, 5, 7, 11, 13 };
3      sl_create(,,, 5,,,, innerprod, sl_glarg(int*,, v1),
4               sl_glarg(int*,, v2), sl_sharg(int, s, 0));
5      sl_sync();
6      printf("%d\n", sl_geta(s)); // prints 143
7      return 0;
8    }
```
Listing 13: Example test program for "innerprod."

As discussed in section 3 and illustrated in listing 5, a thread family can define a daisy-chained communication pattern between all threads. Listing 13 demonstrates a potential use of the thread function "innerprod." This program defines two vectors, then creates a family of 5 threads running "innerprod." It then sends the vector base addresses via the first two channel endpoints, left unnamed, and the value 0 as source value for the "shared" (daisy-chained) third channel, labeled "s." After synchronization on termination, it uses `sl_geta` to read the value sent to the "shared" channel by the last thread.

```
1    sl_seta(x, 3); // invalid: x written to before sl_create
2    sl_create(,,,,,,, foo, sl_glarg(int, x));
3    int y = sl_geta(x); // invalid: x read from before sl_sync
4    sl_sync();
```
Listing 14: Example invalid use of channel endpoints.

Note that the source channel endpoints are not defined prior to `sl_create`, and the final endpoints of "shared" channels cannot be used with `sl_geta` before `sl_sync`. For example, the two uses of `sl_seta` and `sl_geta` in listing 14 are invalid.

Finally, it is also possible to create a family without synchronizing on its termination, i.e. inform the underlying SVP platform that the creating thread

```
1  #include <stdio.h>
2  sl_def(progress) { printf("computing...\n"); } sl_enddef
3
4  int main(void) {
5    int v1[5] = { 1, 2, 3, 4, 5 }, v2[5] = { 3, 5, 7, 11, 13 };
6
7    sl_create(,,,,,,, progress); sl_detach(); // asynchronous
8
9    sl_create(,,, 5,,,, innerprod, sl_glarg(int*,, v1),
10             sl_glarg(int*,, v2), sl_sharg(int, s, 0));
11   sl_sync();
12   printf("%d\n", sl_geta(s)); // prints 143
13   return 0;
14 }
```

Listing 15: Example use of `sl_detach` to define an asynchronous task.

and the created family can proceed asynchronously. This is done by using the word "`sl_detach`" instead of `sl_sync`. For example, the program in listing 15 creates such a *detached* family of one thread to print a progress message, so that the creating thread can immediately start computing while the operating system displays the message towards the user. Note that this program contains a race condition, where the final result could be printed before the text "computing..." We revisit this in the following sections.

For completeness, we detail the grammar of thread function declarations and definitions in fig. 2. This mentions two extra syntax elements "*placement*" and "*create specifier*" which we also detail in the following sections.

### 4.3 Resources and placement

As introduced in section 3, every family of logical threads executes within a *context*, which is a specific set of hardware threads and/or cores on chip. This context is *allocated* prior to family execution and must be *released* (deallocated) afterwards. The SL construct `sl_create`...`sl_sync` automatically combines context allocation, family creation, family synchronization and context de-allocation, i.e. it does not let the program code control allocation and deallocation explicitly. However, the program code can specify *where* on chip the context must be allocated, using the *placement* parameter to `sl_create`.

Any given placement contains two components: a *location* on chip, and a *size* which specifies how many cores to use from that location. By default, the following holds:

- any independent family (with no channels or only "global" channels) is spread automatically using an even distribution over its placement; for example with $N$ logical threads and a placement size $P$, each core at the location will run approximately $N/P$ logical threads;
- any dependent family (with at least one "shared" channel) will only run on the first core of its placement, regardless of the placement size.

When the placement is left unspecified in `sl_create`, it defaults to 0. The value 0 is special and specifies that the context must be allocated at the *same resource* as the family running the creating thread, i.e. that both location and

*c-block-item:*
    ...
    *create-construct*
*create-construct:*
    `sl_create(`, *create-params* , *tfun-exp* [ , *chan-spec* ]* `);` ¬
        *c-block-items* ¬
        *sync-part* ;
*create-params:*
    [ *placement* ] , [ *start* ] , [ *limit* ] , [ *step* ] , [ *wsize* ] , ¬
        [ *create-specifier* ]
*chan-spec:*
    `sl_glarg(` *endpoint-def* `)`
    `sl_sharg(` *endpoint-def* `)`
    `sl_glfarg(` *endpoint-def* `)`
    `sl_shfarg(` *endpoint-def* `)`
*endpoint-def:*
    *c-declaration-specifiers* , [ *identifier* ] [ , *c-assignment-expression* ]
*sync-part:*
    `sl_sync()`
    `sl_detach()`
*create-specifier:*
    `sl__exclusive`
    `sl__forceseq`
    `sl__forcewait`

Figure 2: BNF grammar for thread family creation in SL.

```
sl_def(bar) { } sl_enddef

void foo(int p) {
  sl_create(, p,,100,,,, bar);
  sl_sync();
}
```

Listing 16: Example use of the "inherit" placement.

size must be *inherited*. Consider for example listing 16. If a thread X running `foo(0)` is created on a cluster of 16 cores starting at core 64, then the family running `bar` created by X will also be created using 16 cores starting at core 64. Each core will subsequently run 6 or 7 logical threads `bar` (100/16). If another thread Y running `foo(0)` is created with size 1 at core 13, then the family running `bar` created by Y will be constrained to also run entirely on core 13, even though it defines 100 logical threads.

The value 1 also has a special meaning. It specifies that the context must be allocated at the *same core* as the creating thread, with *size 1*. For example, if a thread X running `foo(1)` happens to execute on core 67, then every thread running `bar` will also run on core 67, regardless of the placement size of the family running X. In other words, a family created with placement 1 becomes *local* to the core running its creating thread. This restriction is further inherited automatically by any descendant family created with placement 0.

In [18, Chap. 11 & App. E] we detail the format of placement addresses and present an additional API to compute placement addresses dynamically:

- `sl_placement_t`: an integer data type large enough to hold placement addresses;
- `sl_default_placement()` evaluates to the explicit placement address of the current family;
- `sl_placement_size(`$P$`)` evaluates to the size of placement address $P$;
- `sl_first_processor_address(`$P$`)` evaluates to the location of placement address $P$;
- `sl_local_processor_address()` evaluates to the location (not size) of the current thread;
- `sl_placement(`$L$`, `$S$`)` evaluates to a placement address for location $L$ and size $S$.

These primitives can be combined to create arbitrary placement addresses. For example, `sl_placement(sl_local_processor_address(), 1)` evaluates to an explicit address equivalent to the special value 1 described above.

The reason why placement is relevant is one of performance control and cost estimation. As suggested in section 5.2, the design of SL discourages programmers (or higher-level code generators) from assuming a minimal amount of parallelism, for instance so that any SL code can run sequentially if no parallelism is available at run-time or if parallelism is disabled. However the very purpose of parallelism is to improve performance, and the *scalability* of a computation depends on the way work is *distributed* over the resources actually available. While distribution is trivial for mostly sequential programs that occasionally invoke operations with a single level of data parallelism, finer-grained control is typically desired when using families of two concurrency levels or more.

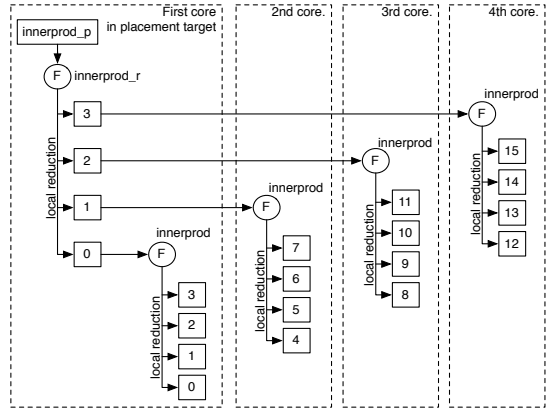

Figure 3: Distribution for listing 17 with n=16 and placement size=4.

Consider for example listing 5. A family running `innerprod` would be dependent and would thus be constrained to run on only one core. To parallelize this function, we can use instead listing 17, deriving listing 5. Here `innerprod_p` first computes how many vector indices to process per core, using `sl_placement_size` to determine the number of cores effectively available. Then it creates a family running `innerprod_r` locally on the first core, one

```
1  int innerprod_p(size_t n, int *a, int *b)
2  {
3    sl_place_t pl = sl_default_placement();
4    int ncores = sl_placement_size(pl);
5    int span = n / ncores;
6
7    pl = sl_first_processor_address(pl);
8
9    sl_create( , /* local placement: */ 1,
10              0, ncores, 1,,, innerprod_r,
11              sl_glarg(int*, , a), sl_glarg(int*, , b),
12              sl_sharg(int, s, 0),
13              sl_glarg(int, , span), sl_glarg(sl_place_t, , pl));
14   sl_sync();
15
16   return sl_geta(s);
17 }
18 sl_def(innerprod_r, , sl_glparm(int*, a), sl_glparm(int*, b),
19         sl_shparm(int, s), sl_glparm(int, span),
20         sl_glparm(sl_place_t, fp))
21 {
22   sl_index(coreid);
23
24   int lower = sl_getp(span) * coreid;
25   int upper = lower + sl_getp(span);
26
27   sl_create(, sl_placement(fp + coreid, 1),
28             lower, upper, 1,,, innerprod,
29             sl_glarg(int*, , sl_getp(a)),
30             sl_glarg(int*, , sl_getp(b)),
31             sl_sharg(int, sr, 0));
32   sl_sync();
33
34   sl_setp(s, sl_geta(sr) + sl_getp(s));
35 }
36 sl_enddef
```

Listing 17: Two-stage reduction, deriving listing 5.

thread per core in the local cluster. Each thread running `innerprod_r` subsequently creates a family running `innerprod` *at* the core for which it has an index. The resulting distribution for n=16 and a starting placement size of 4 cores is given in fig. 3. Note how the code is fully independent of the actual placement size and would run just as well with predictable speedup on either 1 or 100 cores.

Note also that the example in listing 5 is simplified for clarity; it does not support cases where `n` is not a multiple of the number of cores.

### 4.4 Mutual exclusion via exclusive contexts

We have seen in listing 15 a program containing a race condition: the "computing..." message may be printed after the final result, depending on the actual scheduling at run-time. One way to avoid this race condition is to wait until the first message is printed before starting the computation or printing the final result. Another way is to request from the underlying platform that the two calls to `printf` are performed *in sequence*, i.e. exclusive to each other and in program order.

For this, we can use a special SVP feature, called *exclusive contexts*: on every core, there is exactly one context that is never used for regular family creations, but that can be targeted with `sl_create` in combination with the specifier `sl__exclusive`. Because this context is singular, any creation request to a core where the exclusive context is currently busy will call the creation to wait until the context becomes available, that is, until the previous family at that context terminates. This way, mutual exclusion is obtained.

```
#include <stdio.h>
sl_def(progress) { printf("computing...\n"); } sl_enddef
sl_def(final, , sl_glparm(int, r))
{ printf("%d\n", sl_getp(r)); }
sl_enddef

int main(void) {
  sl_create(,,,,,, sl__exclusive, progress); sl_detach();

  /* ... work ... */

  sl_create(,,,,,, sl__exclusive, final, sl_glarg(int,,r));
  sl_detach();

  return 0;
}
```
Listing 18: Example use of mutually exclusive detached families.

Consider for example listing 18. Here `main` creates a first asynchronous family running `progress` at the default exclusive context. The print of "computing..." thus occurs concurrently with the work in `main`. When the work has completed, `main` creates another asynchronous family at the same exclusive context to print the result. Although `main` does not synchronize on termination of either family, the order of the outputs is preserved by mutual exclusion and creation order.

Note that this mechanism is orthogonal to, and combines with the placement information described in the previous section. Namely, each core in the SVP platform has its own exclusive context; two families created concurrently to the exclusive context of *different cores* will not be mutually exclusive. In other words, the combination of `sl_create` and `sl__exclusive` implements *named critical sections*, where each available core constitutes a different section name.

### 4.5 Finite resources and automatic serialization

An SVP platform, including the Microgrid, implements a finite number of hardware threads per core and may not offer preemption to time-share hardware threads between families. This implies that a core can *run out of contexts* to serve family creation requests.

Conversely, for any given use of `sl_create` in a thread, at the point the construct is reached during execution there may or may not be a context available at the target placement to serve the request. Two behaviors can be envisioned: either *wait until a context becomes available*, or *avoid creating a parallel family and instead let the creating thread run the corresponding workload sequentially, in its own hardware context*.

By default, regular creations will not wait and automatically serialize the workload if the target placement has no available context. This mechanism guarantees progress of execution for all family creations; it is described in more detail in [18, Chap. 10]. When so desired, a program can override this behavior per instance of `sl_create` as follows:

- when `sl__forcewait` is specified, the creation will wait until some context can be allocated at the target placement, no matter what. This implies the possibility for deadlock, but may be desirable for system code.
- when `sl__forceseq` is specified, the creation is always serialized in the creating thread, even if some context is available at the target placement. This feature is intended for testing or comparing performance between parallel and sequential execution.

.

Meanwhile, any use of `sl__exclusive` implies `sl__forcewait` implicitly and this cannot be overridden. The corollary is that any work sent to a placement address with `sl__exclusive` must terminate within a finite amount of time, lest it will deadlock any other user of the same exclusive context.

# 5 Background and design rationale

## 5.1 Language primitives vs. an API

Once new features are introduced in a machine interface, two ways exist to exploit them from programs: *encapsulation* in APIs and *embedding* into the language via new primitive constructs recognized by compilers and associated new translation rules to machine code[3].

Encapsulation is technically trivial, and it is desirable when porting existing software using established API such as the POSIX thread interface. However the following must be considered on the Microgrid architecture, which implements fine-grained thread synchronization. As explained in [18, Chap. 4], a thread that issues a long-latency asynchronous operation, e.g. memory load or thread creation, uses regular ISA register names for the endpoints of the communication channels with the asynchronous operation. Meanwhile, the Microgrid hardware protocols can interleave the asynchronous thread management operations (e.g. "allocate," "create") or inter-thread communication operations with other instructions *from the same thread*. To exploit this opportunity with encapsulation, the two phases must be part of separate API functions, and code generators must be configured to avoid reusing these ISA register names for computations while a thread management or communication operation is ongoing. Otherwise, any register spills between phases will cause the thread to wait prematurely for completion of the operation and waste an opportunity for overlapping instruction execution with the asynchronous operation. The same applies for synchronizers that hold the future of asynchronous completions: if the synchronizer that holds a future is spilled, this would cause the creating thread to wait prematurely on the asynchronous operation. To address these issues, a code generator would need to perform an inter-procedural register allocation; furthermore, if the API implementation is compiled separately from the application code, register allocation must then be deferred until all objects are available. Given that no publicly available compiler framework had support for link-time inter-procedural register allocation prior to our work, encapsulation seemed impractical and embedding remained as the unavoidable strategy.

---

[3]In the case of C, extensions via the preprocessor's `#pragma` feature is a form of embedding as it influences code generation.

That said, there is also a quantitative reason as to why embedding is more desirable. The Microgrid architecture allows fine-grained, short-latency thread management and inter-thread communication; and the cost of diverting the control flow for a procedure call is large compared to the synchronization latency (e.g. 40 processor cycles to transfer control to a different procedure vs. 6 cycles to create a family of one thread and 0-1 cycle to communicate a scalar value from one thread to another). In this circumstances, the choice to embed the SVP primitives as new C language primitives reduces the overhead to exploit the synchronization and scheduling granularity offered by the Microgrid.

## 5.2 General design directions

Our original design motivation was to entice code generators and programmers to *expose the fine-grained concurrency* of numerical computations, even the partial concurrency available in dependent computations, in order to enable the automated mapping in hardware of all program fragments, even a few instruction long, to separate cores or hardware threads on the Microgrid architecture.

Another motivation was to *promote resource-agnosticism*, that is promote the expression of programs in a style where the semantics stay unchanged should the hardware parameters evolve. In particular, the approach should discourage programmers from assuming, or knowing, or restricting at run-time the specific amount of effective parallelism (e.g. the number of processors or thread contexts available) when constructing algorithms. This is because otherwise the program is tailored to a specific hardware topology and must be redesigned upon future increases of parallelism. This requires language mechanisms that can express concurrency mostly via *data dependencies* and *declarative concurrency*, in disfavor of explicit control of individual thread creation and placement, and explicit inter-thread communication. When these features are used, it becomes possible to scale the run-time performance of a program by changing the amount of parallelism, and without changing the machine-level encoding of the program. Conversely, it also becomes possible to run any concurrent program on a single processor, since the program cannot assume a minimal amount of parallelism. Thus the flexibility required for dynamically heterogeneous systems [18, Chap. 2] is achieved. We detail this second objective further in [22]; they are shared with other language designs, such as Cilk [6] or more recently Chapel [9].

Furthermore, we decided to place an extra requirement on our language extensions: ensure that the C compiler can generate valid *sequential* code for any new language constructs towards existing (legacy) downstream tools and architectures, e.g. commodity desktop computers. This enables proper troubleshooting and analysis of behavior using standard debugging tools. This is akin to requiring Cilk's *faithfulness* [19] or Chapel's *serializability* [10].

### 5.3 Practicality matters

The inception of SL occurred at a point in time where a large research project around the Microgrid architecture was stalled due to the lack of a usable interface language. While compatibility with other SVP platforms was also considered while defining the language, the SL constructs were first and foremost designed to ensure the short-term availability of a compilation path towards the Microgrid platform.

This "bootstrapping" process and corresponding technical hacks are further detailed in [18, App. H]. To summarize, there was no time to invest in modifying an existing C compiler. Instead, we found a way to instrument the program source code and "trick" the GNU C Compiler (GCC) into generating nearly valid Microgrid assembly code. The instrumentation is performed by translating the SL constructs with a text pre-processor, i.e. in a *context-free* manner, to plain C code using GCC extensions. The use of context-free preprocessing in turn implied the following:

- all the SL constructs must be recognized (tokenize) without considering the surrounding C program text. This in turn favored a uniform syntactic form: a keyword followed by an opening parenthesis, followed by comma-separated extra constructs (which may include balanced inner parentheses), followed by a closing parenthesis;
- since C expression types cannot be decided in a context-free manner at the point of use, all channel uses in the `sl_create` construct must explicitly specify the communication data types;
- as some instrumentation is necessary at the end of a thread program body, and a context-free analysis cannot match block braces recursively, an extra word (`sl_enddef`) is necessary at the end of a thread function.

Code instrumentation occurs *after the regular C preprocessing* and *before C compilation* proper. We use M4 to translate the C-preprocessed SL code to a structured abstract syntax tree, which is then loaded into a program translator which performs the instrumentation, generates plain C code with GCC extensions, calls GCC "behind the scenes" to compile the result, and then post-processes the generated assembly. The full compilation process is captured behind a single compiler driver, called `slc`, which we document more fully in [18, Chap. 6 & App. H].

### 5.4 Implementation status and audience

This implementation-driven design was successful at enabling rapid prototyping and refining of the SL tool chain. We could prototype a first working compiler within one man-month; we later added support for four additional Microgrid targets and two conventional POSIX back-ends within an approximate budget of twelve man-months; this compares favorably with the multiple man-years typically necessary to extend a mature C compiler to a new platform.

SL is currently used as the intermediate representation for most Microgrid-related research at the University of Amsterdam and technology partners. Prior to the publication of [18], multiple published works have pretended to use $\mu$TC for evaluation whereas their results were actually based on the SL technology [3, 14, 13, 26, 23, 21, 28]; this substitution was judged acceptable by their authors due to the resemblance between the two languages. SL has also been targeted successfully by the SAC2C compiler for the array functional language Single-Assignment C [12, 27], as well as a parallelizing C compiler [25, 26].

This article pertains to version 3.7 of the SL tool chain. This tool chain has been released an open source license; its source code, installable packages and related documentations can be accessed from the SVP web site[4].

# 6 Future work

As of this writing, further research and development on the SL language is focused on four areas:
- *software support*: the amount of standard C library services available in SL depends on the underlying SVP platform. For example, as of version 3.7 of the tool chain only a small subset of the POSIX functions can be used in SL code on Microgrid platforms. Future work is planned to extend this support, including implementing a subset of the POSIX threading API so that existing multithreaded code can be run without changes;
- *memory-based communication*: the original design of SL assumes a shared memory, where all threads observe a consistent view of the shared memory's contents. However, as discussed in [18, Chap. 7] and [30], this assumption may not hold in future many-core chips where memory is likely to be disconnected and require explicit communication to move data from one part of the chip to another. Some work has already been started [20] to add partial support for distributed memory to SL, however this still needs to be fully integrated;
- *new communication and synchronization patterns*: the current restriction of channel connections to only the "global" and "shared" patterns, described in section 3, as well as the constraint that synchronization on termination can only be performed in the same thread that also creates a family, makes SL too limited to support general patterns of concurrency. As we discuss in [18, Sec. 6.3.5 & Chap. 12], both SVP and SL should be extended to express other communication and synchronization patterns;
- *abstraction of common distribution patterns*: in particular, distributed reductions and automated load balancing currently require "boilerplate code" that could be captured behind simple, reusable language constructs. Future work is planned to analyze common code patterns and extend the language accordingly.

Meanwhile, the SVP framework itself is an ongoing research project. Any further improvement or development to SVP will likely propagate to SL.

---

[4]Currently `http://www.svp-home.org/`.

# 7 Summary and conclusions

The SL language is a combination of the standard C language with new constructs to define thread functions (`sl_def`) and create families of logical threads running these thread functions (`sl_create`). Threads can communicate using named channels defined and connected together in a manner similar to regular C function parameters and arguments.

Creation is a parameterized construct which captures three dimensions of concurrency. One is *bulk creation*, that is the ability to define a configurable number of threads using a single construct. One is *distribution*, that is the ability to define how threads are spread over hardware resources. The last is *usage policy*, that is the ability to determine behavior upon resource exhaustion or sharing. This latter dimension is in turn exploited (not to say "abused") to obtain *mutual exclusion* as a side feature.

The language was designed with expediency in mind: instead of placing focus on user friendliness, SL was designed towards simplicity of implementation. In particular, it avoids analyzing the surrounding C program text and is thus *unaware of C data types*. As a consequence, SL features *manifest types* for the creation construct where a type-aware implementation would be able to derive types automatically. It also features *separate keywords for integer and floating-point scalars*, which would be otherwise unnecessary for a type-aware implementation. These features were judged an acceptable cost for the relative ease of porting the SL tool chain to new target SVP platforms.

## Acknowledgements

This research was supported by the European Union under grant numbers FP7-215216 (Apple-CORE) and FP7-248828 (ADVANCE). The author would like to thank Andrei Matei for his valuable comments and contributions to the design of SL, as well as Carl Joslin, Irfan Uddin, Michiel W. van Tol, Mike Lankamp, Qiang Yang, Stephan Herhut and Jian Fu for their thorough evaluation and testing of the SL tool chain.